\documentclass[journal]{IEEEtran}
\usepackage{latexsym}

\usepackage{graphicx}
\usepackage{amsfonts,amssymb,amsmath}
\usepackage{cite}
\def\BibTeX{{\rm B\kern-.05em{\sc i\kern-.025em b}\kern-.08em T\kern-.1667em\lower.7ex\hbox{E}\kern-.125emX}}

\newtheorem{lemma}{Lemma}
\newtheorem{prop}{Proposition}

\newcommand{\bfx}{\mathbf{x}}
\newcommand{\bfy}{\mathbf{y}}

\newcommand{\bba}{{{\bf a}}}
\newcommand{\bbx}{{{\bf x}}}

\newcommand{\bfz}{\mathbf{z}}
\newcommand{\bfn}{\mathbf{n}}
\newcommand{\bfa}{\mathbf{a}}
\newcommand{\bfd}{\mathbf{d}}
\newcommand{\bfs}{\mathbf{s}}

\newcommand{\dfn}{\stackrel{\triangle}{=}}

\begin{document}

\title{On Information Rates over a Binary-Input Filtered Gaussian Channel}

\author{M. Peleg, T. Michaeli, and S. Shamai (Shitz)~\IEEEmembership{Life-Fellow, IEEE}
\thanks{The work of S. Shamai has been supported by the European Union's Horizon 2020 Research and Innovation Programme, Grant Agreement
No.~694630, and partly by the WIN Consortium via the Israel Minister of Economy and Science. The manuscript
was submitted to the IEEE Open Journal of the Communications Society (OJ-COMS).}
\thanks{Michael Peleg is with the Viterbi Faculty of Electrical
and Computer Engineering, Technion---Israel Institute of Technology, and also with RAFAEL,
e-mail: peleg.michael@gmail.com.}
\thanks{Tomer Michaeli and Shlomo Shamai (Shitz) are with the Viterbi Faculty of Electrical
and Computer Engineering, Technion---Israel Institute of Technology,
e-mails: tomer.m@ee.technion.ac.il, sshlomo@ee.technion.ac.il.}}

\maketitle

\begin{abstract}
We study communication systems over band-limited Additive White Gaussian Noise (AWGN) channels in which the transmitter's output is
constrained to be symmetric binary (bipolar).
We improve the available
Ozarow-Wyner-Ziv (OWZ) lower bound on capacity which is based on peak-power constrained pulse-amplitude modulation, by introducing
new schemes (achievability) with two advantages over the studied OWZ schemes.
Our schemes achieve a moderately improved information rate and they do so with much fewer sign
transitions of the binary signal. The gap between the known upper bound, which is based on spectral constrains of bipolar signals, and our new
achievable lower bound is reduced to 0.93~bits per Nyquist interval at high SNR.
\end{abstract}


\section{Introduction and Problem Definition}
\label{sec-1}

\begin{figure}[h!t!b!]
\centering
\scalebox{.7}{\includegraphics{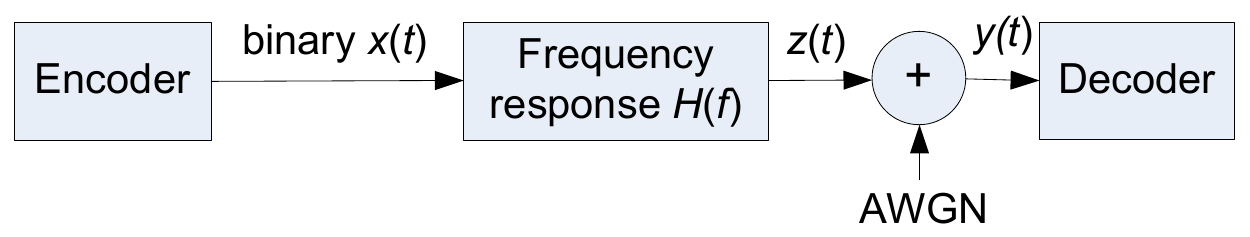}}
\caption{Communication system with binary-valued transmitted signal.}\label{Figure_1}
\end{figure}

\IEEEPARstart{W}{e} study communication systems over band-limited Additive White Gaussian Noise (AWGN)
channels in which the transmitter's output is constrained to be bipolar, as presented in Figure~\ref{Figure_1}.
Such systems arise when the power efficiency must be high or when the transmitter needs to be of very low complexity.
Those systems are usually implemented by some form of Pulse Width Modulation (PWM), Pulse Position Modulation (PPM), or similar schemes,
operating over Gaussian noise channels \cite{1-hpc},\cite{2-frg},\cite{3-paa}.
Communication systems with binary transmitted signals are of recent practical interest in millimeter-wave wide-band applications,
e.g. \cite{4-lmsy}, \cite{5-ndf}.

In this work, we examine theoretical limits on communication with binary transmission, not limited to PWM.
We are interested in the reliable information rate supported by this system focusing mainly on the region of
asymptotically high $\mathrm{SNR}$. This theoretical problem was addressed by Ozarow, Wyner, Ziv (OWZ) \cite{6-owz} using the
Pulse Amplitude Modulation (PAM) method. OWZ \cite{6-owz} showed that performance, measured by mutual information,
achievable with a signal peak-limited to $\pm \sqrt{P}$ can also be achieved with a binary-valued $\pm \sqrt{P}$ signal with a very high Sign
Transition Rate (STR). They applied this finding to design a PAM scheme with symbols uniformly distributed in
$  \bigl[-\sqrt{P} , + \sqrt{P} \bigr] $, which provides an achievable lower bound on the capacity of the system.
As implied by \cite{6-owz}, peak-limited continuous-time signals such as filtered PAM, can in
principle be also band limited \cite{7-shamai88} and hence represented by sampling at an appropriate rate, while the equivalent
(in the sense of \cite{6-owz}) bipolar processes cannot be strictly bandlimited \cite{8-shepp}.
A lower bound exceeding for low $\mathrm{SNR}$ that of \cite{6-owz}, was presented in \cite{9-sow}, based on improved bounds for
intersymbol-interference Gaussian channels. Additional results on capacity of systems with binary
inputs, some with additional constraints on average transition rate, minimum inter-transition time and out of band power, are
presented in \cite{10-shamai93} and \cite{11-sbd}. Systems with limited minimal transition times were investigated in \cite{12-cs},
including systems with mild filtering, that is, not strictly bandlimited as in \cite{6-owz}.

The binary channel input carries information in its transition times. Sampling the binary input at a Nyquist rate
corresponding to the channel bandwidth would degrade the performance severely, thus the system in Figure~\ref{Figure_1}
falls in general into the category of Faster Than Nyquist (FTN) signaling which is of wide current theoretical
and practical interest. In recent years FTN signaling approaches, forms and extensions of classical pulse-amplitude modulation
strategies have emerged. See overviews of these relevant domains in \cite{13-fgzrlc},\cite{14-aro},\cite{15-zqycwgt} and references therein.
See also recent examples of advanced theory and techniques in \cite{16-kcz},\cite{17-lybb},\cite{18-gzj},\cite{19-kbv}.
FTN can provide significant advantages in terms of capacity with prescribed modulation techniques and signaling strategies,
though the resultant channel may suffer significant inter-symbol-interference, which demands higher complexity detection
procedures. Yet, no peak-power restrictions are imposed on the resultant time-continuous process,
which is a central part in our scheme. This is motivated by practical constraints, as was also the case in \cite{6-owz},
reflecting the constraints of magnetic storage media.

In this work we present new schemes with two advantages over \cite{6-owz}.
They achieve a moderately improved information rate and do so with much fewer sign transitions
of the binary signal. The new schemes require STR of only up to twice the Nyquist rate of the channel, while \cite{6-owz} uses STR many folds higher
than the Nyquist rate; if implemented fully, the STR in \cite{6-owz} is infinite. Low STR is easier to implement in systems which are already wide-band
and in which each sign transition must pass a power amplifier such as \cite{4-lmsy}, \cite{5-ndf}.
We extended the technique in \cite{6-owz} to the new schemes in which the transmitted signal is a non-linear function of the information sequence.
The studied communication system is presented in Figure~\ref{Figure_1}.
\newline
It comprises an encoder producing a binary-valued $\pm \sqrt{P}$ input $x(t)$ where $P$ is the transmit power,
AWGN channel with noise Power Spectral Density (PSD) of $ \frac{1}{2}\, N_0 $ watt/Hz (double-sided) and a receiver.
The channel has a frequency response $ H(f) $, in our case a unity frequency response at frequencies from $0$ to $B$
and zero otherwise.  The channel output $y$ is
\begin{equation}
\label{eq-1}
y(t) = z(t) +n(t) \, ,
\end{equation}
where $z(t)$ is the filtered desired signal and $ n(t)$ is the Gaussian noise.

We denote by $B$ the bandwidth of the low pass brick-wall filter in Hz, $T= 1/(2B) $ is the Nyquist sampling period
associated with $B$, $~ \rho =\frac{P}{N_0 B} $ is the signal to noise ratio $\mathrm{(SNR)}$, $\log$ denotes the
natural logarithm and bold lower-case letters denote vectors and sequences.

\section{Known Performance Bounds}
\label{sec-2}

Shamai and Bar-David \cite{20-sbd}, derived an upper bound on the system capacity, based on the fact that the Power Spectral Density (PSD)
of a binary-valued signal is limited by certain constraints presented in \cite{21-McMillan} and \cite{8-shepp}.
They analyzed limits on spectral densities of binary signals and then
upper bounded the capacity of the system by Mutual Information (MUI) when the channel input has the capacity achieving Gaussian
distribution with the same PSD as the binary-valued signal. For high $\mathrm{SNR}$ they proved that relative to
the capacity-achieving frequency-flat Gaussian input there is a power loss at least by a factor of $\gamma= 0.9337$,
see definition of $\gamma$ below.
The same paper considers Random Telegraph Signal (RTS) as an interesting example rather than a bound and the power factor
there is around $\gamma= 0.63 $ which is an upper bound on the performance of RTS.
The capacity $ C_G $ bits/second of the channel with PSD given as $ S(f) $  used in \cite{20-sbd}
is the well-known expression
\begin{equation*}
C_G = \int\limits_0^B\, \log_2  \, \left(1+\displaystyle\frac{S(f)}{N_0}\right) \, df \, ,
\end{equation*}
where $C_G$ is achieved with a Gaussian input. In the limit of asymptotically high $ \frac{S(f)}{N_0} $ the capacity $ C_G $
becomes
\begin{equation}
\label{eq-2}
C_G^h = \int\limits_0^B\, \log_2  \,\displaystyle\frac{S(f)}{N_0} \, df \, .
\end{equation}
For a frequency-flat Gaussian signal of bandwidth $B$ this yields
\begin{equation*}
C_{G0}^h = B\cdot \log_2  \,\displaystyle\frac{P}{N_0 B} \, .
\end{equation*}
Multiplying $S(f) $ in \eqref{eq-2} by a factor $\gamma$ increases $C_G^h $ by $B \cdot \log_2 \,  \gamma $ information bits per
second which are $ \Delta = \frac{1}{2}\, \log_2 \, \gamma$ bits per Nyquist sampling interval.
Consequently, the equivalent $\mathrm{SNR}$ gain is defined,
for a scheme with bandwidth~$B$, as a function of difference $\Delta$ in information per Nyquist interval between
the scheme and the AWGN channel with the same bandwidth and transmit power as,
\begin{equation*}
\gamma = 2^{2\Delta} \, .
\end{equation*}
OWZ \cite{6-owz} derived the following achievable lower bound using the modulation method \cite{6-owz} described in the introduction.
\begin{equation}
\label{eq-3}
I_{\mathrm{OWZ}} \ge \displaystyle\frac{1}{2} \,\log_2 \, \left(\displaystyle\frac{2P\cdot e}{\pi^3 N_0B} +1\right)
 = \displaystyle\frac{1}{2} \,\log_2 \, \left( \rho \,\displaystyle\frac{2e}{\pi^3} +1\right)
\end{equation}
where $ I_{\mathrm{OWZ}} $ stands for mutual information per Nyquist interval.

This corresponds to
\begin{equation*}
\gamma_{\mbox{\tiny OWZ}} = \displaystyle\frac{2e}{\pi^3} = 0.1753 \, .
\end{equation*}
The bipolar signal that achieves the performance of
the PAM modulation technique in \cite{6-owz} involves high transition
rate of the binary signal. An improved lower bound in the low $\mathrm{SNR}$ regime is reported in \cite{9-sow}.

\section{New Achievable Schemes}
\label{sec-3}

The main results of this work are the improved lower bounds on the capacity of the bipolar-input bandlimited AWGN channel,
see Proposition~\ref{prop-1}. The proposition is proved by introducing and analyzing new communication schemes.

We discuss four schemes, denoted by $ \mathrm{A}, \mathrm{B}, \mathrm{B1} $ and $ \mathrm{C}$.
In all of them the time axis is partitioned into successive intervals of
duration $T$ equal to the Nyquist interval corresponding to $\mathit{B} $.
In scheme $\mathrm{A} $, the binary signal in each interval $n$ of time $t$ spanning $ (n-0.5) T\leq t < (n+0.5) T $ is
\begin{equation}
\label{eq-4}
x(t) = \begin{cases}
~~1 & \quad (n- 0.5) T \leq  t\leq (n+a_n)T\\
-1 & \quad (n+ a_n)T < t < (n+ 0.5)T
\end{cases} \, ,
\end{equation}
where $a_n$ are the information-carrying variables, uniformly, independently and identically distributed
(u.i.i.d.) over $ [-0.5 , +0.5] $. Thus, information is conveyed by the time of sign reversal of the signal,
see Figure~\ref{fig2} for an illustration.

\begin{figure}[h!]
\centering
\scalebox{.7}{\includegraphics{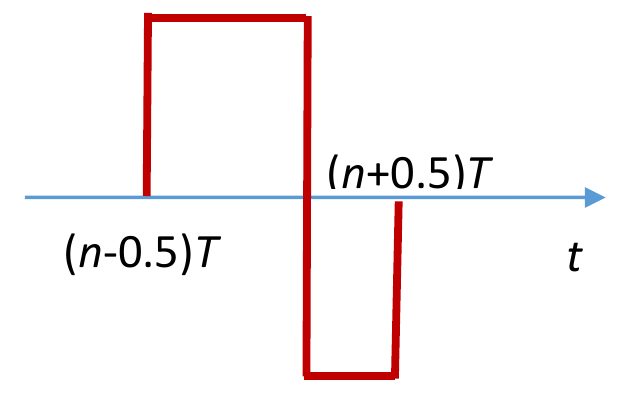}}
\caption{Binary signals $ x(t) $ of type~$ \mathrm{A} $.
The single symbol with a variable transition time.}\label{Figure_2}
\end{figure}

We denote the sequence of all $a_n$ by $\bba $, denote the binary transmitted signal in interval $n$ as
$ x_n (t) $ and over all the transmission by $x(t)$ or $\bbx$.

\begin{figure}[h!]
\centering
\scalebox{.5}{\includegraphics{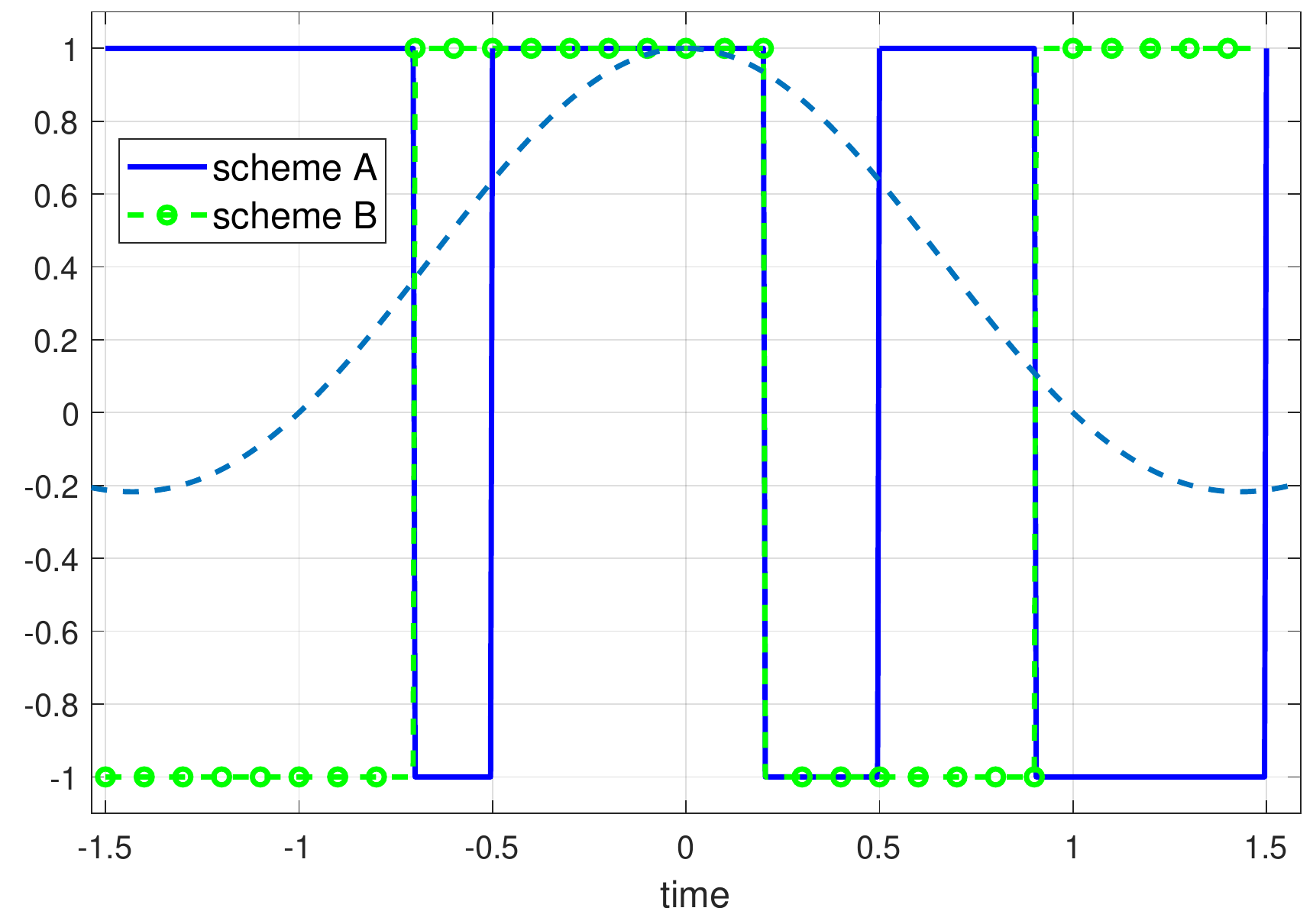}}
\caption{Binary signals $ x(t) $ of type~$ \mathrm{A} $ and $ \mathrm{B} $.
The dashed line is the impulse response of the channel filter.}\label{Figure_3}
\end{figure}

Scheme $ \mathrm{B} $ is derived from scheme $ \mathrm{A} $ by inverting the signal in successive intervals of length $T$
to eliminate half of the sign transitions of the binary signal. See Figure~\ref{Figure_3}.

Scheme $ \mathrm{C} $ is derived from scheme $ \mathrm{A} $ by inverting the signal in successive intervals at random
where the signs $ s_n $ valued as $\pm 1 $ are used as additional information inputs. The signs $ s_n $ are equi-probable
and independent. The signaling in scheme~$ \mathrm{C} $ comprises $ a_n $ and $ s_n $, thus the signaling rate is twice
the Nyquist rate. Scheme~$ \mathrm{B1} $ is introduced below. The STR of schemes~$ \mathrm{A} $, $ \mathrm{B}, \mathrm{B1} $ and
$ \mathrm{C} $ is $4B$, $\mathrm{2B} $, $2B$ and $3B$ correspondingly by construction, see  Figure~\ref{Figure_3}, while the Nyquist rate is $2B$.
Denote by $\bfs$ the sequence of the sign inversions $s_n$ in schemes~$ \mathrm{A} $, $ \mathrm{B} $ and
$ \mathrm{C} $, so that $ s_n = -1 $ for the inverted symbols and $s_n = 1 $ otherwise.

Computing the exact capacity of the three schemes, that is, the MUI between the binary input $x(t)$ and
the channel output $y(t)$, seems intractable. We therefore we resorted to deriving upper and lower bounds.
\begin{figure}[h!]
\centering
\scalebox{.5}{\includegraphics{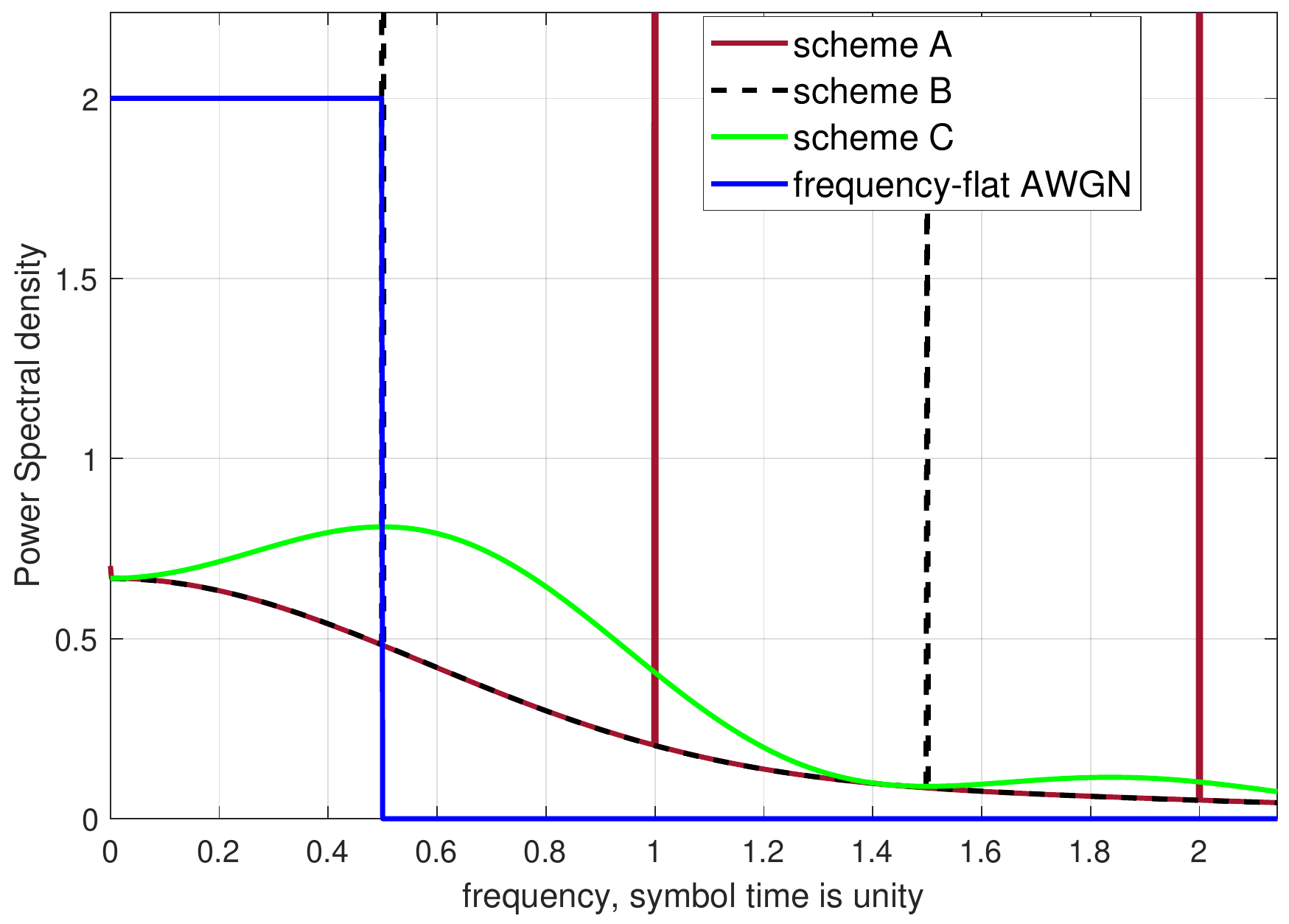}}
\caption{Spectrum of the three schemes and of the frequency-flat AWGN.
The continuous parts of the scheme~$ \mathrm{A} $ and scheme~$ \mathrm{B} $ curves overlap.}\label{Figure_4}
\end{figure}

To compute upper bounds on the communication rates of our schemes, we first evaluate the PSD of the signals.
We assume that the signal is randomly shifted as a whole by a delay distributed uniformly over $ (0,T) $ to render it stationary.
The autocorrelations of the signals in the three schemes are derived in the appendix and summarized in (5).
\begin{figure*}[t!]
\begin{eqnarray*}
& R_A (\tau)  = \begin{cases}
\left(1-\displaystyle\frac{2|\tau |}{T}\right) \left(1-\displaystyle\frac{|\tau |}{T}\right) +
\left(-\displaystyle\frac{2}{3}\left|\displaystyle\frac{\tau}{T}\right|^3 + 2\left|\displaystyle\frac{\tau}{T}\right|^2 -
\left|\displaystyle\frac{\tau}{T}\right|\right) & ; \quad \displaystyle\frac{|\tau|}{T} < 1\\
\displaystyle\frac{1}{3} + 2\tau_n^2 - 2\tau_n & ; \quad  \mbox{otherwise}
\end{cases} \hfill {\mbox{(5a)}} \\
\noalign{\noindent\mbox{where} $\tau_n = \displaystyle\frac{|\tau |_{\mbox{mod} \, T}}{T} \, $ \mbox{and}
$ \, |\tau |_{\mbox{mod} \, T} \, $  \mbox{denotes the modulo}  $T$ \mbox{operation.}}\\
& R_B (\tau)  = \begin{cases}
\left(1-\displaystyle\frac{2|\tau |}{T}\right) \left(1-\displaystyle\frac{|\tau |}{T}\right) -
\left(-\displaystyle\frac{2}{3}\left|\displaystyle\frac{\tau}{T}\right|^3 +
2\left|\displaystyle\frac{\tau}{T}\right|^2 - \left|\displaystyle\frac{\tau}{T}\right|\right) & ; \quad
\displaystyle\frac{|\tau|}{T} < 1 \\
\displaystyle\frac{1}{3} (4\tau_n^3 - 6\tau_n +1) \cdot (-1)^{\lfloor |\frac{\tau}{T} |\rfloor} & ;
\quad \mbox{otherwise}
\end{cases} \hfill {\mbox{(5b)}} \\
\noalign{\noindent\mbox{where} $\lfloor \tau \rfloor \, $  \mbox{is the largest integer smaller than} $ \,\tau$}
\nonumber \\
& R_C (\tau)  =
\begin{cases}
\left(1-\displaystyle\frac{2|\tau |}{T}\right) \left(1-\displaystyle\frac{|\tau |}{T}\right) & \; ;  \quad
\displaystyle\frac{|\tau|}{T} < 1\\
0 &  \; ; \; \mbox{otherwise}
\end{cases} \hspace*{4.8cm} {\mbox{(5c)}}
\end{eqnarray*}
\hrulefill
\end{figure*}

The PSD was obtained by numerical Fourier transform of the autocorrelations, see Figure~\ref{Figure_4},  and verified by
simulation. The PSD is obtainable analytically from the autocorrelations. For example for scheme $ C $ with $T=1$
we have the one-sided PSD:
\begin{equation*}
S_c (f) = \displaystyle\frac{3\pi f - 2\sin (2\pi f) + \pi f \cos (2\pi f)}{(\pi f)^3} \, .
\end{equation*}

The AWGN line in Figure~\ref{Figure_4} is the PSD of the standard bandlimited capacity-achieving signal without the binary constraint.
As well-known from the water-pouring theory, it spreads the available power uniformly over the available bandwidth~$B$.
The PSDs of our three schemes suffer the disadvantage of wasting some of the transmitted power out of the channel
bandwidth and of not spreading the remaining power uniformly.
Scheme~$ \mathrm{C} $ is evidently better than schemes~$ \mathrm{A} $ and $ \mathrm{B} $.
Indeed the schemes~$ \mathrm{A} $, $ \mathrm{B} $ and $ \mathrm{C} $ are constrained to bipolar transmitted signals
and therefore cannot possess a strictly bandlimited spectrum, as we know from \cite{8-shepp}.
The spectra of schemes A and B are identical except for the discrete
frequency components (tones) which do not influence the outcome of \eqref{eq-2}.
There are no discrete tones in scheme~$ \mathrm{C} $ since it decorrelates the pulses by random sign inversions,
limiting the support of the autocorrelation to $[-T, T] $.

Based on the PSD, we compute the upper bounds on performance at high $\mathrm{SNR}$ of the three schemes using
\eqref{eq-2} and compare them to the optimal input which is a Gaussian signal with power~$P$ and a flat PSD from~$0$ to
$ \mathit{B} $. The results are presented in Table~\ref{tab1}.
{\small
\begin{table}[h!]
\fontsize{10}{8}
\renewcommand{\arraystretch}{1.1}
\caption{Upper bounds using Gaussian inputs
with the same spectra}
\begin{center}
\begin{tabular}{|l|c|c|} \hline
& Equivalent power & Information loss \\
& factor $\small{\gamma}$ & $ \small{\Delta} $ relative to\\
Scheme & relative to & rectangular spectra\\
& rectangular & 0 to $B$ Hz in\\
& spectra 0 to $B$ Hz & bits/Nyquist interval\\\hline
Rectangular spectra & & \\
0 to $B$ Hz & 1 & 0\\\hline
Schemes A and B & 0.3 & 0.8686\\\hline
Scheme C & 0.367 & 0.7232\\\hline
Upper bound & & \\
on binary & 0.9337 & 0.0495\\
schemes in [20] &&\\\hline
Upper bound on & & \\
the Random & 0.6271 & 0.3366 \\
Telegraph signal as in [20] &&\\ \hline
\end{tabular}
\label{tab1}
\end{center}
\end{table}
}

We proceed to derive lower bounds on communication rates of the new schemes.
As shown in Figure~\ref{Figure_1}, $x (t) $ passes through the channel filter and is then contaminated by AWGN.
The receiver filters the signal by the same low pass filter, which is clearly an information-lossless operation.
We sample the filtered channel output at the Nyquist rate $1/T$ producing an infinite
sequence $\bfy$ of samples $y_n$. We denote the signal without the noise component by a sequence $\bfz$ of samples
$z_n$, see Figure~\ref{Figure_1}.

We lower-bound the capacity $I(\bfx ; \bfy) = H(\bfy)-H(\bfy |\bfx ) $ by adapting the approach presented in OWZ \cite{6-owz}.
Since $ H(\bfy|\bfx) $ is the known entropy of the noise, the main term to evaluate is $ H(\bfy) $.
OWZ lower-bounded $ H(\bfy) $ as a function of the entropies of its components $ H(\bfz) $ and $ H(\bfn) $
using the Entropy-Power Inequality (EPI) presented in \cite{22-blahut}. OWZ evaluated $ H(\bfz) $ using the fact that the channel was an Inter
Symbol Interference (ISI) channel representable by a Toeplitz matrix
the determinant of which is computable using the Szeg\"{o} theorem \cite{24-gs}.

We begin by determining the entropy of $ \bfz $. The required differential entropy is
\begin{equation*}
h_z = \displaystyle\frac{1}{N}\,  h(z_1 \, \dotsc z_N) \, .
\end{equation*}
In schemes~$ \mathrm{A} $ and $ \mathrm{B} $, each $a_n$ determines one symbol $x_n$ and those symbols are linearly
filtered to produce $\bfz$. The sequence $\bfa $, treated as a vector in the next equation, comprises u.i.i.d.  components,
and therefore its differential entropy is:
\setcounter{equation}{5}
\begin{equation}
\label{eq-6new}
h_a \dfn \displaystyle\frac{1}{N}\, h(\bfa ) = h(a_i) = \log (1) \, .
\end{equation}

The noiseless sampled output $\bfz$ is a function of $\bfa$, which we denote by $\bfz=m(\bfa)$.
To derive $ h(\bfz) $ using the Jacobian formula \eqref{eq-6}  similarly to \cite{6-owz},
we need our transformation $ \bfz = m(\bfa) $ to be a bijection
and $ \bfa $ and $\bfz$ must have identical dimensions.
\begin{lemma}
For every $ \varepsilon > 0 $, if the channel's bandwidth is $ \mathit{B} = \frac{1}{2T} + \varepsilon$,
where $ T $ is the signaling period, then the transformation $ \bfz = m(\bfa ) $ in schemes~$ \mathrm{A} $ and
$ \mathrm{B} $ is a bijection.
\end{lemma}

\begin{IEEEproof}
The modulation scheme in Figure~\ref{Figure_1} is deterministic, therefore each sequence $\bfa $ can produce only a single
sequence $\bfz$.  It remains to prove that there are no two distinct sequences $\bfa$ producing the same $\bfz$.
If this would happen, then there would exist a pair of transmitted signals $ \bfx_1 \neq \bfx_2 $ such that
$ \bfz (\bfx_1 )= \bfz (\bfx_2 ) $, implying  $ \bfd \overset{\mbox{\scriptsize{def}}}{=}  \bfz (\bfx_1) - \bfz (\bfx_2 ) = 0 $.
Since the low-pass filter is linear, such a $\bfd$ would be the low-pass filtered signal $ \bfx_1 - \bfx_2 $.
By the construction of $\bfx$, for schemes~$ \mathrm{A} $ and $ \mathrm{B} $, not $ \mathrm{C} $,
the difference $ \bfx_1 - \bfx_2 $ would be a sequence of pulses as depicted in Figure~\ref{Figure_5} in which
each pulse is assigned a symbol interval $T$ during which it has a zero value except for some contiguous duration in
which it is $ \pm 2 $ , see Figure~\ref{Figure_5}.
\begin{figure}[h!]
\centering
\scalebox{.7}{\includegraphics{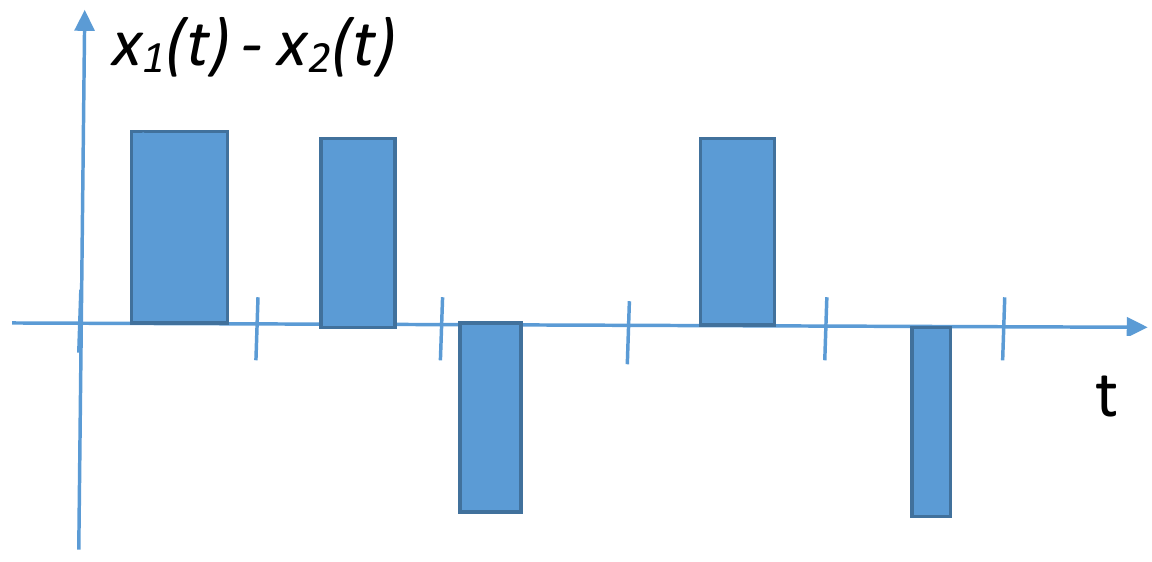}}
\caption{Difference signal $ \bfx_1 - \bfx_2 $.}\label{Figure_5}
\end{figure}

So it is sufficient to prove that such a nonzero signal $\bfx_1 -\bfx_2 $ cannot have zero spectra in $ 0\le f\le B + \epsilon $.
This follows directly from \cite[Theorem 1]{23-en}, which proved that signals with zero spectra in $ 0\le f\le B + \varepsilon $,
which are denoted in \cite{23-en} as high-pass signals or signals with a zero gap, change sign at average rates higher than
$ 1/T=2\, B $, which is the highest possible rate of sign changes of the function $ \bfx_1 -\bfx_2 = \bfx_1 (t) - \bfx_2 (t) $
in Figure~\ref{Figure_5}. Thus, such a nonzero $\bfx_1 -\bfx_2 $ cannot exist and $ \bfz = m(\bfa) $ is a
bijection. The asymptotically small change in $B$ is immaterial in this work by the problem definition.
\end{IEEEproof}

Since $ \bfz = m ( \bfa ) $ is a bijection, the entropy $ h_z $ is
\begin{align}
\label{eq-6}
h_z \dfn \displaystyle\frac{1}{N}\, h(\bfz) = & \displaystyle\frac{1}{N}\, h(\bfa) + \displaystyle\frac{1}{N}\, \int\, p(\bfa) \,
\log \left|\displaystyle\frac{\partial z_i}{\partial a_j}\right| \, d\bfa \nonumber\\
= & \displaystyle\frac{1}{N}\, h(\bfa) + \displaystyle\frac{1}{N}\, E_{\bfa} \,
\left( \log \,\left|\displaystyle\frac{\partial z_i}{\partial a_j}\right|\right)  \, ,
\end{align}
where $ \Bigl|\frac{\partial z_i}{\partial a_j}\Bigr| $ denotes the determinant of the Jacobian matrix
of $ \bfz = m (\bfa ) $, $ p(\bfa) $ is the probability density function of $\bfa$ and $ E_{\bfa} $
denotes expectation with respect to $\bfa$.
The Jacobian matrix is denoted $ \Bigl(\frac{\partial z_i}{\partial a_j}\Bigr) \dfn J $.
Unlike OWZ \cite{6-owz}, in our scheme ~$ \mathrm{A} $, the Jacobian matrix is not Toeplitz since here
$ \frac{\partial z_i}{\partial a_j} $ depends on each $ a_j $.
Therefore, we could not follow OWZ using the Szeg\"{o} theorem \cite{24-gs}.
Instead, we evaluated the expectation in \eqref{eq-6} numerically by generating the signals $ \bfz $ with random sequences $ \bfa $,
computing $J$ for each $ \bfz $ and averaging $ h_z $. The Jacobian matrix $J$ is evaluated by
\begin{equation}
\label{eq-6newest}
\displaystyle\frac{\partial z_i}{\partial a_j} = \pm 2\,
\displaystyle\frac{\sin \, (\pi t_{ij}/T)}{\pi t_{ij}/T} \, ,
\end{equation}
where $ t_{ij} $ is the time elapsed from the time of transition $ a_i $ to the sample $ z_i $.
The sign is positive for transitions from $1$ to $-1$ and negative otherwise.
The computation was executed on cyclic sequences 500 and 1000~symbols long and verifying identical result in both cases.

Denote
\begin{equation*}
h_d = h_z -h_a = \displaystyle\frac{1}{N}\,\displaystyle\int\, p(\bfa) \log \,
\left| \displaystyle\frac{\partial z_i}{\partial a_j} \right| \,  d\bfa \, .
\end{equation*}
The result of numerical evaluation is $h_d = 0.5197 $~nats for $ T=1 $ and $P=1$
and is invariant with $T$, see \eqref{eq-6new} and \eqref{eq-6newest}.

The entropy \eqref{eq-6} is identical in scheme~$ \mathrm{A} $ and in scheme~$ \mathrm{B} $ with its alternate sign inversions.
This is because $ \frac{\partial z_i}{\partial a_j} $ changes sign when $ \bfs_j = -1$,
so for scheme~$ \mathrm{B} $ we can create a new auxiliary vector
\begin{equation}
\label{eq-7}
\hat{\bfa} = (a_1 s_1 , \,\dotsc \, a_i s_i \, , \dotsc )
\end{equation}
in which $ \Bigl(\frac{\partial z_i}{\partial \hat{a}_j}\Bigr) $ is identical to $ \Bigl(\frac{\partial z_i}{\partial a_j}\Bigr) $
in scheme~$ \mathrm{A} $ and $ h(\hat{\bfa}) = h(\bfa) $ yielding the same $ h_z $
in schemes~$\mathrm{A} $ and $ \mathrm{B} $.

The true entropy of $z$ is larger by $ 0.5 \log (P) $ due to multiplication by $ \sqrt{P} $ and, $a$ has a unity support, so
$ h_a = 0 $, see \eqref{eq-6new}. Thus,
$$
h_z = \displaystyle\frac{1}{N}\, h(\bfz ) = h_d + 0.5 \log P \, .
$$

The entropy of the sampled noise at filter output is
$$
h_n = 0.5 \log \, (2\pi e BN_0 ) \, .
$$
By EPI \cite{22-blahut}, the entropy of the sum is upper bounded in terms of entropies of its components:
\begin{align*}
e^{2h_y} & \ge e^{2h_z} + e^{2h_n}\\
e^{2h_y} & \ge e^{2(h_d + 0.5 \log (P))} + 2\pi e BN_0\\
h_y & \ge 0.5 \log \, \Bigl(e^{2h_d + 0.5 \log (P)} + 2\pi e BN_0 \Bigr)
\end{align*}
\setcounter{equation}{8}
\begin{align}
\label{eq-9new}
I_{ya} & \dfn\displaystyle\frac{1}{N}\, I(\bfy ; \bfa) = h_y - h_{y|\bfa}\\
& \ge 0.5 \log \Bigl( e^{2h_d + \log (P)} + 2\pi e B N_0 \Bigr) - 0.5 \log (2\pi e BN_0 ) \nonumber
\end{align}
\begin{align}
\label{eq-10new}
I_{ya} & \ge 0.5 \log \, \left(\displaystyle\frac{e^{2h_d + \log (P)} + 2\pi e BN_0}{2\pi eBN_0}\right) \nonumber\\
I_{ya} & \ge 0.5 \log \, \left(\displaystyle\frac{\displaystyle\frac{1}{2\pi e}\, P \cdot e^{2h_d} + BN_0}{BN_0}\right)
\end{align}
The AWGN capacity is
$$
C_{\mathrm{AWGN}} = 0.5 \log \, \left(\displaystyle\frac{P+BN_0}{BN_0}\right) \, .
$$
So the power gain at all $ \mathrm{SNR} $s over the AWGN channel is lower-bounded by
$$
\gamma = \displaystyle\frac{1}{2\pi e} \, e^{2h_d} = 0.1656 \, .
$$
OWZ \cite{6-owz} reported a better result, $ \gamma_{\mathrm{OWZ}} = 0.1753 $, see \eqref{eq-3} above;
our software reconstructs this result as a verification.

The same analysis technique used here for the brickwall channel response is applicable to a general channel frequency response $H(f)$.
To extend the technique to a more general $ H(f) $, the sinc pulse used in \eqref{eq-6newest} to compute the Jacobian matrix and shown in
Figure~\ref{Figure_3}
would be replaced by the new channel impulse
response. Furthermore, $H(f)$ would need to be non-zero over $ 0 <f < B $ to fulfill the conditions of Lemma~1.

Next we show an improved performance in scheme~$ \mathrm{C} $. To increase $h(z)$, the polarity of each pulse is inverted at random.
As seen in Figure~\ref{Figure_4} this also removes the wasted discrete tones from the signal spectra.
The analysis above cannot be applied directly since now $  x(t) \to z(t) $
is not a bijection as demonstrated by construction of pairs of signals $x(t)$ the difference of which have period of $T$ and a zero mean,
thus zero PSD in the $0$ to $ \mathrm{B} $ frequency band. For scheme~$ \mathrm{C} $ the system mutual information between
the modulator inputs $\bfa , \bfs $ and the channel output is:
\begin{equation}
\label{eq-9}
I (\bfa , \bfs ; \bfy ) = I (\bfs ; \bfy ) + I(\bfa ; \bfy | \bfs ) \, .
\end{equation}
The second term on the r.h.s. is equal to schemes~$ \mathrm{A} $ and $ \mathrm{B} $, the signs $\bfs$ on which this term
is conditioned are treated by the auxiliary vector $ \hat{\bfa} $ as defined in \eqref{eq-7} for scheme~$ \mathrm{B} $.
The first term on the r.h.s. is the improvement achieved by scheme~$ \mathrm{C} $ relative to schemes~$ \mathrm{A} $
and $ \mathrm{B} $. We lower-bound it as follows. Denote the sequence of derivatives of $y(t)$ at times $nT$ by
$ \dot{\bfy} = \{ \dot{y}_n\} $.

\begin{figure}[h!t!b!]
\centering
\scalebox{.5}{\includegraphics{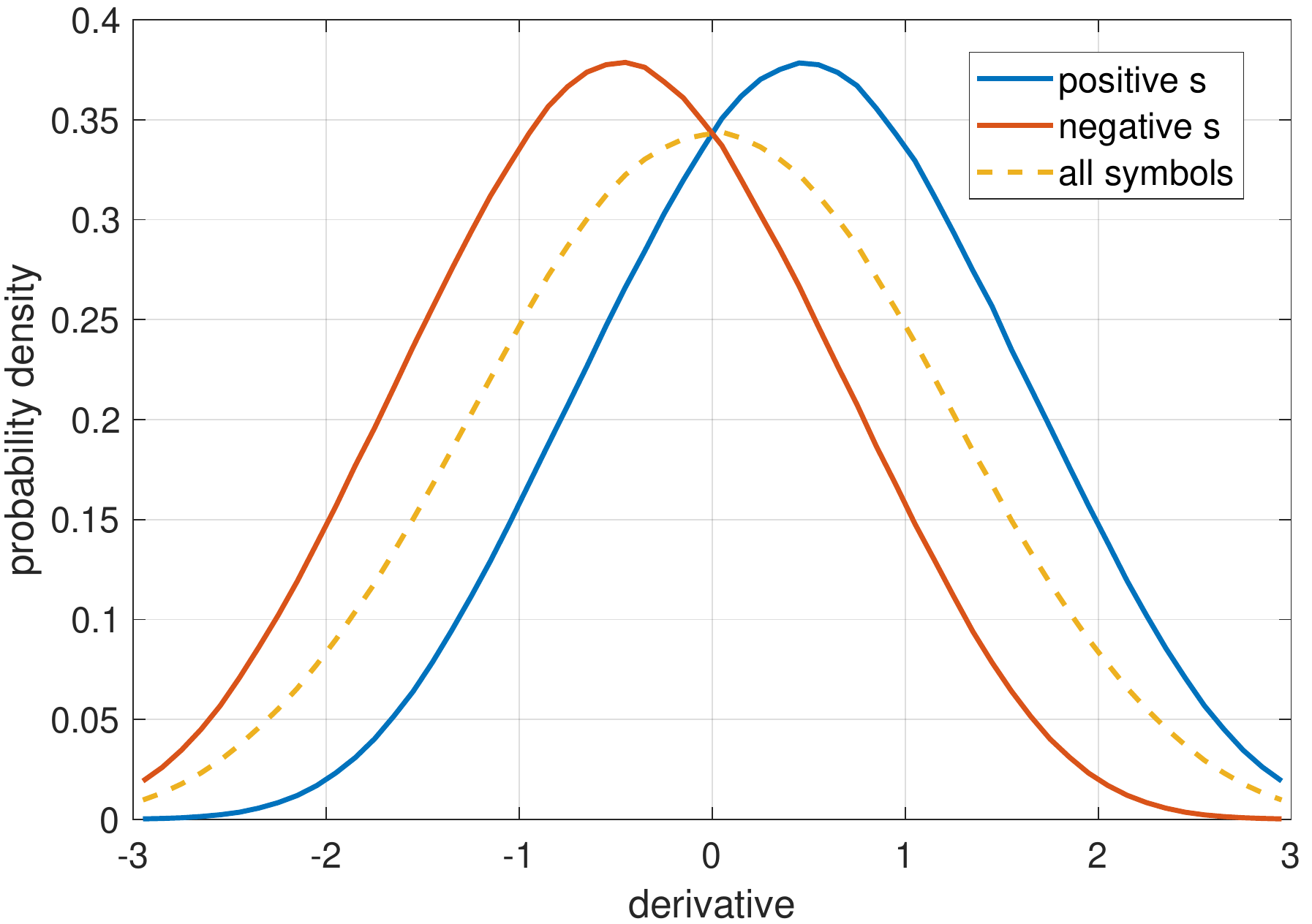}}
\caption{Probability density functions of signal derivatives conditioned on signs $ s_n$.}\label{Figure_6}
\end{figure}

Now
\begin{align}
I(\bfs ; \bfy ) & \geq I (\bfs ; \dot{\bfy}) \nonumber\\
& = \sum\limits_n \, I(s_n ; \dot{\bfy} | \bfs_1^{n-1}) \, ; \qquad \bfs_1^{n-1} = s_1 \, \cdots \, s_{n-1}\nonumber\\
& \geq \sum\limits_n\, I(s_n ; \dot{\bfy})\nonumber\\
& \geq \sum\limits_n\, I(s_n ; \dot{y}_i) \, .
\end{align}
The first line is since $ \dot{\bfy} $ is a function of $ \bfy $.
The second line is by the standard mutual information decomposition \cite{22-blahut}.
The third line is since $s_n$ is independent of $\mathbf{s_1^{n-1}} $.
The last term was evaluated by simulation of scheme~$ \mathrm{C} $ while
estimating the symbol-wise probability densities $ P(\dot{y}_n | s_n = 1 ) $, $P(\dot{y}_n | s_n = -1) $ and $ P(\dot{y}_n ) $ as plotted in
Figure~\ref{Figure_6}.
It adds 0.136~bits per symbol at asymptotically high $ \mathrm{SNR} $ which is equivalent to a power gain of $\gamma = 1.207$.
Scheme~$ \mathrm{C} $ achieves $\gamma= 0.20$, moderately better than OWZ.
We expect that better detectors would improve upon this lower bound.

\begin{figure}[h!]
\centering
\scalebox{.7}{\includegraphics{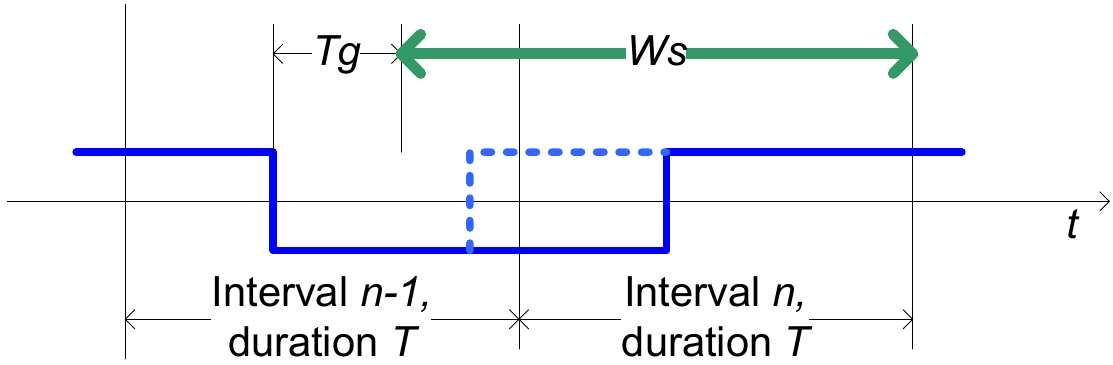}}
\caption{Time diagram of scheme~$ \mathrm{B1} $, the solid blue line is the transmitted signal, the dotted blue line is another possible transmitted
signal.}\label{Figure_7}
\end{figure}

We noticed that each pair of consecutive sign-transitions with exceptionally short inter-transition time introduces a very low singular value to $J$ which
reduces our lower bound. To address this we designed scheme~$ \mathrm{B1} $.
Scheme~$ \mathrm{B1} $ improves upon scheme~$ \mathrm{B} $ by introducing a minimal inter-transition interval $ T_g = 0.2 T $ and by extending
the range in which each sign-transition time can occur. In particular, as in scheme~$ \mathrm{B} $, each transmission interval of duration $T$
is associated with one sign-transition. However, the transition time specified in \eqref{eq-4} as uniformly distributed over the $n$'th
transmission interval spanning $ (n - 0.5)T \leq t < (n + 0.5) T $ in scheme~$ \mathrm{B} $, is, in the new $ \mathrm{B1} $ scheme,
distributed uniformly over a window $ W_s $, see Figure~\ref{Figure_7}, which starts $ T_g $ after the
previous sign-transition and ends, as in scheme~$ \mathrm{B} $, at the end of the current interval.
This is illustrated in Figure~\ref{Figure_7}. Note that the sign transition
associated with the $n$'th transmission interval may occur in the $n$'th interval or in one of the few intervals preceding it.

Lemma~1 holds also for scheme~$ \mathrm{B1} $ in which the difference signal is as in Figure~\ref{Figure_5} with the same average number of pulses except
for not confining each pulse to its own $T$-interval. That is, in both the schemes~$ \mathrm{B} $ and $ \mathrm{B1} $, the total number of all
the negative and positive pulses in the difference signal is half of the total number of sign transitions in $ x_1 $ and $ x_2 $.

The entropy of $ \bfa $  in \eqref{eq-6new} is now calculated numerically as
\begin{align}
h_a \dfn & \displaystyle\frac{1}{N}\, \sum\limits_n h(a_n | a_ 1^{n-1}) \nonumber\\
= & \displaystyle\frac{1}{N}\, \sum\limits_n \, \log \Bigl( W_s (n)\Bigr) \, .
\end{align}
It is larger by 0.4095 nats than that of scheme~$ \mathrm{B} $, contributing to the performance. Scheme~$ \mathrm{B1} $ achieved the best performance
among the four schemes, see Table~\ref{tab2}.

With  scheme~$ \mathrm{B1} $, the achievable lower bound has an advantage of a power factor of 1.47 at all $ \mathrm{SNR} $s and of
0.28~bits per Nyquist interval $T$ at high $ \mathrm{SNR} $ over the scheme reported
in \cite{6-owz}. Comparing to Table~\ref{tab1}, the gap between the upper and the lower bounds specific to the schemes is 0.43 and 0.438~bits
per Nyquist interval for schemes~$ \mathrm{A} $ and $ \mathrm{C} $ respectively.
The gap between the upper bound in \cite{20-sbd}, entry~6 in Table~\ref{tab2}, and the best achievable lower bound, entry~4 in the table, is
0.93~bits per Nyquist interval.
{\small
\begin{table}[h!t!b!]
\renewcommand{\arraystretch}{1.1}
\fontsize{10}{8}
\caption{Comparison of different approaches}
\begin{center}
\begin{tabular}{|l|c|c|} \hline
Scheme & Equivalent power & Information loss at \\
& factor $\small{\gamma}$ & high SNR relative to\\
& relative to & Gaussian signal with\\
& Gaussian signal with & rectangular\\
& rectangular & spectra 0 to $B$ Hz \\
& spectra 0 to $B$ Hz & in bits/Nyquist interval\\\hline
Gaussian signal with & & \\
rectangular spectra & 1 & 0\\
0 to $B$ Hz &&\\\hline
OWZ , [6], achievable & &\\
lower bound & 0.1753 & 1.256\\\hline
Schemes A and B, &  & \\
achievable& 0.1656 & 1.2973\\
lower bound &&\\\hline
Scheme B1, achievable & 0.2586 & 0.976\\
lower bound &&\\\hline
Scheme C, achievable & 0.20 & 1.1613\\
lower bound &&\\\hline
Upper bound in [20] & 0.9337 & 0.0495\\\hline
Upper bound on RTS & 0.6271 & 0.3366\\
as in [20] &&\\\hline
\end{tabular}
\label{tab2}
\end{center}
\end{table}
}

The three schemes have distinct attributes.
Schemes~$ \mathrm{A} $ and $ \mathrm{B} $ serve to build up the theoretical base and they provide a lower bound on capacity
valid for all $ \mathrm{SNR} $s. Scheme~$\mathrm{C} $ is an extension providing an improved lower bound at high $ \mathrm{SNR} $
and an improved spectra. Scheme~$\mathrm{B1} $ provides the best lower bound at all $ \mathrm{SNR} $s.

\begin{prop}
\label{prop-1}
If the capacity of the AWGN frequency-flat low-pass channel with a given average input power serves as a baseline,
then imposing an additional constraint of a symmetric binary bipolar input does not degrade the capacity more than by a power loss of 0.2586
at all $ \mathrm{SNR}$s and information loss of 0.93~bits per Nyquist interval at high $ \mathrm{SNR} $.
\end{prop}

\begin{IEEEproof}
Compare scheme~$ \mathrm{B1} $ in Table~\ref{tab2} to the upper bound in \cite{20-sbd}.
\end{IEEEproof}

\section{Conclusion and Outlook}

We studied communication systems over the band-limited AWGN channel in which the transmitter output is constrained to be binary
bipolar. We presented new schemes which provide an improved lower bound on the capacity of this channel. The gap between the known
upper bound and our new achievable lower bound is reduced to 0.93~bits per Nyquist interval at high $ \mathrm{SNR} $.
Furthermore, the schemes operate at a much lower rate of sign transitions than the bipolar signaling
that achieves the PAM based bounds in \cite{6-owz}.

There is a room for future work attempting to improve the achievable lower bound. For this purpose signals with spectra more concentrated
in the lower frequency regions than our scheme~$ \mathrm{C} $, see Figure~\ref{Figure_4}, should be investigated.
Interestingly, the maximal power factor $\gamma$ of the
Random Telegraph Signal (RTS) is achieved with average transition rate of about 0.67 per Nyquist interval, less than the 1.5 average
transition rate of our scheme~$ \mathrm{C} $ leading to a narrower PSD, thus a future analysis of performance of the RTS signaling
might reduce the gap between the upper and lower bounds further.

The lower bound on performance presented here might be improved in future work based on techniques that consider PWM and also RTS in terms of lower
bounding the filtered minimum mean square error, and incorporating the Information Estimation relations \cite{27-gsv}.
Further interesting useful techniques developed for ISI channels \cite{25-sl},\cite{26-cs} should also be considered.

In this paper the signals are designed for good performance in the high $ \mathrm{SNR} $ regime while the results for
schemes~$ \mathrm{A} $, $ \mathrm{B} $ and $ \mathrm{B1} $
hold for all $ \mathrm{SNR} $s, see~\eqref{eq-10new}.
Future work may address the non-asymtotic low and intermediate $ \mathrm{SNR} $ region based on new schemes adapted to $ \mathrm{SNR} $
and on advanced FTN techniques listed in the introduction for which the Shamai-Ozarow-Wyner \cite{9-sow} bound is of direct relevance.

\section*{Appendix---autocorrelations}

Denote the autocorrelation of $ x(t) $ as
$$
R(\tau ) = E_{x,t} \Bigl[x(t)\cdot x(t+\tau ) \Bigr] \, ,
$$
where $ E_{x,t} $ denotes expectation over $x$ and over $ -T < t<T $.

For scheme~$ \mathrm{C} $ we have
\begin{equation}
\label{eq-10}
R_c =\begin{cases}
\left(1-\displaystyle\frac{2|\tau |}{T}\right) \left(1-\displaystyle\frac{|\tau |}{T}\right) & \; ;  \quad \displaystyle\frac{|\tau|}{T} < 1\\
0 &  \; ; \; \mbox{otherwise} \; .
\end{cases}
\end{equation}
The first parenthesis is the correlation given that $t$ and $+t$ are in the same symbol interval, the second parenthesis is
the probability of this occurrence. The expression for cases~$ \mathrm{A} $ and ~$ \mathrm{B} $ is a little more involved.
For $ \frac{|\tau |}{T} > 1, x(t) $,  and $ x(t+\tau ) $ are independent, thus
\begin{equation}
\label{eq-11}
R(\tau ) = E_t \Bigl\{ E_x \Bigl[ x(t)\Bigr]\cdot E_x \Bigl[x(t+\tau )\Bigr]\Bigr\} ; \; \frac{|\tau |}{T} > 1 \, ;
\end{equation}
For scheme~$ \mathrm{A} $, $ E\Bigl[x(t)\Bigr] = 1-2/T $. It follows by a straightforward integration for scheme~$ \mathrm{A} $:
\begin{equation*}
R_A (\tau) = \displaystyle\frac{1}{3} + 2\tau_n^2 - 2\tau_n ; \quad \frac{|\tau |}{T} > 1 \, ,
\end{equation*}
where $ \tau_n = \frac{|\tau |_{\mbox{mod}} \, T}{T} $ and $ |\tau |_{\mathrm{mod} \, T} $ denotes the modulo $T$ operation.
For scheme~$ \mathrm{B} $:
\begin{equation*}
R_B (\tau) = \displaystyle\frac{1}{3} (4\tau_n^3 - 6\tau_n^2 +1) \cdot
(-1)^{\lfloor | \frac{\tau}{T} |\rfloor} ; \quad  \frac{|\tau |}{T} > 1 \, ,
\end{equation*}
where $ \lfloor \tau \rfloor $ is the largest integer smaller than $\tau$.

For $ \frac{|\tau |}{T} < 1 $, the expectation over $t$ is the sum over the events in which $t$ and $ \tau +t $ are in the same
symbol interval which yields \eqref{eq-10} and of a term contributed by the events
where $t$ and $ \tau +t$ fall into successive symbol intervals where \eqref{eq-11} applies. The result is:

\bigskip
\begin{center}
$$
\begin{array}{ll}
R_{\mathrm{AB}} (\tau ) &  =
\left(1-\displaystyle\frac{2|\tau |}{T}\right) \left(1-\displaystyle\frac{|\tau |}{T}\right)\\
& \pm \Bigl(-\frac{2}{3}\left|\displaystyle\frac{\tau}{T}\right|^3 + 2\left|\displaystyle\frac{\tau}{T}\right|^2 -
\left|\displaystyle\frac{\tau}{T}\right|\Bigr) \; ; \displaystyle\frac{|\tau|}{T} < 1
\end{array} \quad \mbox{(12)}
$$
\end{center}

\begin{figure}[h!]
\scalebox{.5}{\includegraphics{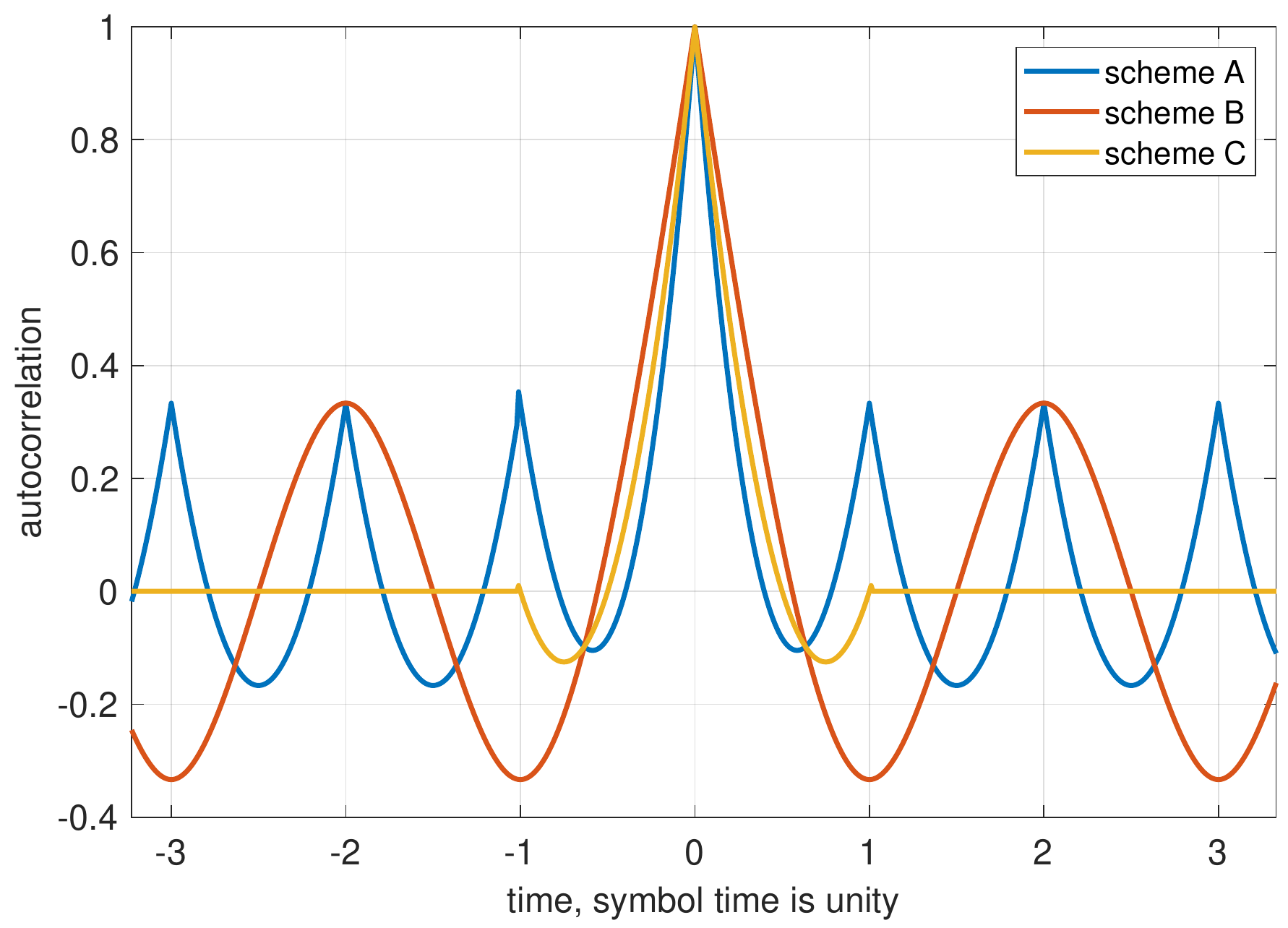}}
\caption{
Autocorrelation functions for schemes~$ \mathrm{A} $, $ \mathrm{B} $ and $ \mathrm{C} $.}\label{Figure_8}
\end{figure}

\medskip
\noindent
where the sign is positive for $ \mathrm{A} $ and negative for $ \mathrm{B} $.
Collecting the equations above yields (5) and Figure~\ref{Figure_8}.

\bibliographystyle{IEEEtran}
\bibliography{tssept21.bib}

\begin{IEEEbiographynophoto}{Michael Peleg}
(M--87, SM--98) (Senior Member, IEEE) received his B.Sc. and M.Sc. degrees from the Technion---Israel Institute of Technology,
Haifa, Israel in 1978 and 1986 respectively.
From 1980 till 2000 he was with the communication research facilities of the Israel Ministry of Defense.
From 2000 till now he is with Rafael Ltd. He is associated with the EE Dept. of the Technion---Israel Institute of Technology,
where he is collaborating in research in communications and information theory.
His research interests include wireless communications, iterative decoding, multi-antenna systems and radiation safety.
\end{IEEEbiographynophoto}

\begin{IEEEbiographynophoto}{Tomer Michaeli}
is an Associate Professor at the ECE faculty of the Technion---Israel Institute of Technology. He completed his B.Sc. and Ph.D degrees
in this faculty in 2005 and 2012, respectively. From 2012 to 2015 he was a postdoctoral fellow at the CS and Applied Math Department at the Weizmann
Institute of Science. His research lies in Computer Vision and Machine Learning. In particular, he studies problems in image restoration, generation,
and manipulation. He won several awards, including the Marr Prize in 2019, and the Krill Prize of the Wolf foundation in 2020.
\end{IEEEbiographynophoto}

\begin{IEEEbiographynophoto}{Shlomo Shamai}
(Life Fellow, IEEE) is currently with the Viterbi Faculty of
Electrical and Computer Engineering, Technion---Israel Institute of Technology, where he is
also a Technion Distinguished Professor, and holds the William Fondiller
Chair of Telecommunications. He is also an URSI Fellow, a Member of the
Israeli Academy of Sciences and Humanities, and a Foreign Member of the
U.S. National Academy of Engineering. He was a recipient of the 2011 Claude
E. Shannon Award, the 2014 Rothschild Prize in Mathematics/Computer
Sciences and Engineering, and the 2017 IEEE Richard W. Hamming Medal.
He was a co-recipient of the 2018 Third Bell Labs Prize for Shaping the
Future of Information and Communications Technology. He was also a
recipient of numerous technical and paper awards and recognitions of the IEEE
(Donald G. Fink Prize Paper Award), Information Theory, Communications
and Signal Processing Societies, and EURASIP. He is listed as a Highly
Cited Researcher (Computer Science) for the years 2004, 2005, 2006, 2007,
2008, and 2013. He has served as an Associate Editor for the Shannon
Theory of the IEEE TRANSACTIONS ON INFORMATION THEORY. He
has also served twice on the Board of Governors for the Information Theory
Society. He has also served on the Executive Editorial Board for the IEEE
TRANSACTIONS ON INFORMATION THEORY, the IEEE Information
Theory Society Nominations and Appointments Committee, and the IEEE
Information Theory Society, Shannon Award Committee.
\end{IEEEbiographynophoto}

\end{document}